# Regional-Scale Estimation of Soil Hydraulic Conductivity Using the Kansas Mesonet

Behzad Ghanbarian[1,2,3*] and Andres Patrignani[4]

[1] iResearchE3 Lab, Department of Earth and Environmental Sciences, University of Texas at Arlington, Arlington 76019 TX, USA

[2] Department of Civil Engineering, University of Texas at Arlington, Arlington TX 76019, USA

[3] Division of Data Science, College of Science, University of Texas at Arlington, Arlington TX 76019, USA

[4] 2004 Throckmorton Hall, Department of Agronomy, Kansas State University, Manhattan, Kansas, USA

[*] Corresponding author's email address: ghanbarianb@uta.edu

**ABSTRACT**

In soil physics, saturated hydraulic conductivity, $K_{sat}$, is among the most important hydraulic properties with broad applications to modeling flow and transport under saturated conditions. Its accurate estimation, however, is challenging and requires precise characterization of pore space. In this study, we applied concepts of critical path analysis (CPA) to estimate $K_{sat}$ from soil water retention curve. To evaluate the CPA, we used 313 undisturbed soil samples from the Kansas Mesonet database in which the value of $K_{sat}$ spans over five orders of magnitude in variation. We found that the CPA estimated $K_{sat}$ reasonably well with root mean square log-transformed error RMSLE = 0.87. For most samples, the predicted values were around the 1:1 line within a factor of 10 of the measurements. We also estimated $K_{sat}$ using five other methods but none was more accurate than the CPA.

## 1. INTRODUCTION

The saturated hydraulic conductivity ($K_{sat}$) of soils is an important hydraulic variable for quantifying water fluxes such as infiltration rate, soil moisture redistribution, and drainage rate under fully saturated conditions modulates. $K_{sat}$ can be measured in field conditions using ring infiltrometers, Guelph permeameters, or disk infiltrometers; in laboratory conditions using open- or close-path permeameters; or can be estimated as a function of other soil properties (e.g., soil texture, porosity) using pedotransfer functions. However, obtaining accurate estimates of $K_{sat}$ of soils has been a longstanding challenge in soil physics because soils can exhibit substantial spatial variability and have complex pore structures due to their particle size distribution, organic matter content, and management that dramatically affect the values of $K_{sat}$. Evidence from the literature indicate that for some soils particle sizes can span more than four orders of magnitude in variation (Wu et al., 1993; Bittelli et al., 1999; Prosperini and Perugini, 2008; Yong et al., 2017). Soils are also deformable porous media (Raats and Klute, 1968) and their structures are dynamic (Sullivan et al., 2022) due to biotic factors like the presence of plant roots and micro-organisms and abiotic factors like the presence of clay minerals that cause soils to shrink and swell with varying levels of soil moisture.

In the literature, numerous relationships including theoretical and empirical models (Chapuis, 2012; Ghanbarian, 2021) as well as pedotransfer functions (Ghanbarian et al., 2015a; Van Looy et al., 2017; Zhang and Schaap, 2019) have been proposed to estimate $K_{sat}$. Among theoretical frameworks, the conceptualization of a soil as a bundle of capillary tubes (Xu and Yu, 2008; Peters et al., 2023), effective medium approximation (Doyen, 1988; Ghanbarian et al., 2019a), renormalization group theory (Karim and Krabbenhoft, 2010; Esmaeilpour et al., 2023), critical path analysis (Skaggs, 2011; Hunt and Sahimi, 2017) and percolation theory (Hunt et al., 2014; Sahimi, 2023) have been applied to model $K_{sat}$ in porous media.

In the soil science community, pedotransfer functions were proposed to estimate $K_{sat}$ from other readily measurable soil properties, allowing researchers to extend the applicability of large, curated databases of soil hydraulic properties (Tietje and Hennings, 1996; Lin et al., 1999; Weber et al., 2024). A prominent example widely used in soil science is Rosetta (Schaap et al., 2001; Zhang and Schaap 2017), which is a machine learning pedotransfer function. Although widely applied, developing accurate pedotransfer functions is challenging. Evidence from the literature indicate that pedotransfer functions are database dependent (Schaap and Leij, 1998; Roustazadeh et al., 2024). This means a pedotransfer function developed and evaluated using a database may fail to accurately estimate a target feature e.g., $K_{sat}$, if assessed further using an independent database including unseen data. Another challenge associated with pedotransfer functions is the number of samples required to train them. The representative number of samples was recently investigated by Ahmadisharaf et al. (2024) who demonstrated that either learning curve or representative sample size analysis is required to determine whether or not the number of samples is enough and to develop reliable regression-based models using machine learning.

In contrast to empirical relationships, such as pedotransfer functions, theoretic models have theoretical foundation, and their parameters are physically meaningful. This means they would provide reasonable estimates of $K_{sat}$, if their parameters are determined accurately. The main objective of this study is to use a database of soil physical properties determined for stations of the Kansas Mesonet to evaluate several theoretical models in the estimation of $K_{sat}$.

## 2. MATERIALS AND METHODS

In this section, we first briefly describe the database of soil physical properties from the Kansas Mesonet and then we present the models from the literature used in this study to estimate $K_{sat}$.

### 2.1. Kansas Mesonet Soil Physical Property Database

The Kansas Mesonet is an environmental monitoring network with nearly 90 stations deployed across the state of Kansas, USA (Patrignani et al., 2020). Each station measures a wide range of atmospheric variables and soil variables like soil temperature and volumetric water content at 5, 10, 20, and 50 cm depth. From 2018 to 2020, about 40 stations were visited and undisturbed soil cores with a volume of 100 cm3 were collected at each sensor depth using a manual auger, totaling 313 intact soil cores (Parker et al., 2022). Each soil sample was saturated from the bottom-up using a 5 mM CaCl₂ solution. Then, $K_{sat}$ was determined using the constant-head method in a closed-path permeameter. For the soils with very low permeabilities, the falling head method was employed in accord with Reynolds and Elrick (2002). All measurements were performed at the controlled temperature of $21 \pm 2°C$.

After measuring $K_{sat}$, the wet-end of the soil water retention curve was measured using a combination of suction table (-2, and -5 kPa) and pressure cells (-10, -33, and –70 kPa). After equilibration at -70 kPa, the soil cores were oven-dried at 105 °C for two days, ground and then sieved using a 2-mm mesh. Then, the water content at a matric potential of -1,500 kPa was determined with a pressure plate extractor. These observations were then used to find the parameters of the van Genuchten soil water retention model (van Genuchten, 1980) using least squares optimization. The van Genuchten model is defined as:

$$\frac{\theta - \theta_r}{\theta_s - \theta_r} = [1 + (\alpha\,\psi_m)^n]^{-(1-1/n)}$$

where $\theta_s$ (cm³ cm⁻³) is the saturated volumetric water content, $\theta_r$ (cm³ cm⁻³) is the residual volumetric water content, and $n$ and $\alpha$ are fitting parameters.

## 2.2. Models of Saturated Hydraulic Conductivity

### 2.2.1 Critical path analysis

Critical path analysis (CPA) is a theoretical technique used to determine the most significant pathways that govern transport properties in disordered systems (Hunt, 2001; Hunt et al., 2014). In porous media, CPA identifies the subset of pore networks that dominate the flow of fluids, effectively simplifying the complex geometry of porous structures into key transport pathways (Hunt and Sahimi, 2017). This method has been successfully applied to model single-phase permeability of rocks (Katz and Thompson, 1986; Ghanbarian et al., 2016; 2021). By considering the heterogeneity and connectivity of pore spaces, CPA enables accurate estimation of permeability from pore-scale properties such as pore throat size distribution and network topology using the following relationship

$$K_{sat} = \frac{\rho g}{\mu} \frac{d_c^2}{C_{CPA} F} \tag{1}$$

where $\rho$ is water density, $g$ is gravitational acceleration, $\mu$ is water viscosity, $d_c$ is the critical pore diameter, $F$ is the formation factor ($=\sigma_w/\sigma_b$, in which $\sigma_w$ is the water electrical conductivity and $\sigma_b$ is the bulk electrical conductivity), and $C_{CPA}$ is a constant coefficient whose value depends on the pore geometry (Ghanbarian et al., 2016). In this study, we set $C_{CPA}$ = 53.5, following Skaggs (2011). The value of $d_c$ was determined from the pore throat diameter at the inflection point of soil water retention curve ($d_{inf}$), following Ghanbarian and Skaggs (2022).

In the Kansas Mesonet database, the electrical conductivity of the soil samples was not measured. We, therefore, estimated the value of $F$ using the Ghanbarian et al. (2014) model i.e.,

$$F = \begin{cases} \frac{(1-\phi_c)(\phi_x-\phi_c)}{(\phi-\phi_c)^2}, & \phi_c < \phi < \phi_x \\ \frac{1-\phi_c}{\phi-\phi_c}, & \phi_x < \phi < 1 \end{cases} \tag{2}$$

where $\phi$ is the porosity, $\phi_c$ is the critical porosity and $\phi_x$ is the crossover porosity. Although Hunt (2004) proposed $\phi_c = 0.1\phi$, in this study we set $\phi_c = \theta_r$. Although the value of $\phi_x$ is non-universal, meaning its value varies from one sample to another (Ghanbarian et al., 2015b), for the sake of simplicity, we set $\phi_x = 0.75$. Ghanbarian and Hunt (2014) showed that $\phi_x = 0.75$ was a reasonable assumption in mineral soils. We should point out that the top equation in Eq. (2) with $\phi_c = 0$ and $\phi_x = 1$ reduces to $F = \phi^{-2}$, known as Archie's law (Archie, 1942).

### 2.2.2 Kozeny-Carman model

The Kozeny-Carman model (Kozeny, 1927; Carman, 1937) has been widely used to estimate $K_{sat}$ of soils (Chapuis and Aubertin, 2003; Regalado et al., 2004; Singh and Wallender, 2008). The model expresses hydraulic conductivity as a function of the square of porosity, divided by tortuosity and specific surface area, with a proportionality constant accounting for fluid properties.

$$K_{sat} = \frac{\rho g}{\mu} \frac{d_a^2 \phi^3}{180(1-\phi)^2} \qquad (3)$$

Despite its simplicity, the Kozeny-Carman model is limited by its inability to account for complex soil heterogeneity. The particle size distribution reported in the Kansas Mesonet database only consists of the percentage for the sand, silt, and clay fractions. Therefore, the value of $d_a$ in Eq. (3) was estimated from the weighted geometric mean of sand, silt and clay sizes using the relationship $d_a = exp\left[\frac{\sum_{i=1}^{3} w_i \ln(d_i)}{\sum_{i=1}^{3} w_i}\right]$ in which $d_i$ represents the average particle size for sand, silt and clay (respectively equal to 0.1025, 0.0026 and 0.0001 cm) and $w_i$ denotes the sand, silt and clay fractions.

### 2.2.3 Glover et al. (2006) model

This model, also known as the RGPZ (Revil, Glover, Pezard and Zamora) model, was analytically derived from electrokinetic theory. It relates permeability to parameters such as mean grain size and pore throat size, offering a more accurate prediction compared to traditional methods. The authors validate their model by comparing its predictions with experimental data, demonstrating improved accuracy over existing models. The RGPZ model is

$$K_{sat} = \frac{\rho g}{\mu} \frac{d_a^2 \phi^{3m}}{4am^2} \qquad (4)$$

where $m$ is the cementation exponent in Archie's law i.e., $F = \phi^{-m}$ (Archie, 1942), and $a$ is a grain shape factor equal to 8/3 for quasi-spherical particles. Similar to the Kozeny-Carman model, we estimated $d_a$ in Eq. (4) from the weighted geometric mean of sand, silt and clay sizes and fractions. Since formation factor was not measured with the Kansas Mesonet database, we set $m = 2$ based on Archie's law. This value is also consistent with the universal scaling exponent in percolation theory (Hunt et al., 2014; Sahimi, 2023).

### 2.2.4 Johnson et al. (1986) model

Johnson et al. (1986) proposed the following theoretical model for saturated hydraulic conductivity estimation from the size of the dynamically connected pores ($\Lambda$) and formation factor ($F$)

$$K_{sat} = \frac{\rho g}{\mu} \frac{\Lambda^2}{8F} \tag{5}$$

Following Ghanbarian et al. (2019b), we set $\Lambda = 0.21 d_{inf}$ and estimated $F$ via Eq. (2) with $\phi_x = 0.75$.

### 2.2.5 Mishra and Parker (1990) model

In conjunction with the van Genuchten (1980) soil water retention curve model, Mishra and Parker (1990) developed the following expression for $K_{sat}$

$$K_{sat} = 108(\theta_s - \theta_r)^{\frac{5}{2}} \alpha^2 \tag{6}$$

where $\theta_s$ and $\theta_r$ are respectively the saturated water content values and $\alpha$ is the van Genuchten model parameter (van Genuchten, 1980). In Eq. (6) the unit of the constant 108 is cm³/s and thus the unit of $K_{sat}$ is cm/s. By comparing with 48 soil samples collected from different locations in Virginia, Mishra and Parker (1990) showed that their model provided reasonable estimations of $K_{sat}$ (see their Fig. 1). However, the soil samples included in the Kansas Mesonet soil physical property database were predominantly classified as fine-textured soils (e.g., silty clay, silty clay loam, clay loam, and silt loam).

### 2.2.6 Guarracino (2007) model

Using concepts of bundle of capillary tubes approach, Guarracino (2007) proposed a model similar to the Mishra and Parker (1990) model as follows:

$$K_{sat} = 4.65 \times 10^4 \phi \alpha^2 \tag{7}$$

where $\phi$ is the porosity and $K_{sat}$ is cm/day. Guarracino (2007) evaluated the predictability of Eq. (7) by comparing it with average values of $K_{sat}$ for 12 soil textural classes reported by Carsel and Paris (1982) and reported reasonable accuracy across observations spanning four orders of magnitude (Fig.1 in Guarracino (2007)).

### 2.3 Goodness of fit

To evaluate the performance of each model we used the root mean squared logarithmic error (RMSLE). The RMSLE was used to evaluate model performance because the measured $K_{sat}$ values span nearly five orders of magnitude. RMSLE quantifies errors in the logarithmic domain, making it more appropriate for variables that vary multiplicatively rather than additively, such as hydraulic conductivity. By assessing deviations on a relative rather than absolute scale, RMSLE assigns comparable weight to under- and overestimations across the full range of $K_{sat}$, thereby reducing the undue influence of extreme values. In contrast, metrics such as the Nash–Sutcliffe Efficiency (NSE) or the root mean square error (RMSE) computed on the original (linear) scale tend to be dominated by high $K_{sat}$ values, leading to biased interpretations of model accuracy. The use of RMSLE thus provides a more balanced and scale-consistent assessment of predictive performance when dealing with data that cover multiple orders of magnitude.

## 3. RESULTS

Results of $K_{sat}$ estimations via different models are presented in Fig. 1. The CPA approach, Eq. (1) in combination with Eq. (2), estimated the $K_{sat}$ with an RMSLE of 0.87 (Fig. 1a). Most predictions lie around the 1:1 line and within a factor of 10 of the measured values (81%). For several samples, however, the estimated $K_{sat}$ values differ from the measured ones by more than one order of magnitude. Since samples were undisturbed, such discrepancies could be attributed to soil structural features – such as root channels, wormholes, or cracks – that enhance macroporosity but are not captured by the soil water retention curve used as input to the CPA. In Eq. (1), we used $C_{CPA} = 53.5$, which resulted in reasonable estimations of $K_{sat}$ for the Kansas Mesonet soils. Ghanbarian et al. (2016) also reported that this coefficient provided accurate predictions in tight gas sandstones. Nevertheless, the fixed value of $C_{CPA}$ introduces a degree of uncertainty, as its optimal magnitude may vary with soil texture, pore structure, and measurement scale.

The Kozeny-Carman model (Eq. 3) tended to overestimate $K_{sat}$ for lower measured values and underestimate it for higher ones, yielding an RMSLE of 1.42 (Fig. 1b). Such discrepancies can be explained by (1) the model limited validity for unconsolidated porous media with narrow particle size distributions (Dullien, 1992) and (2) uncertainties in estimating $d_a$ from the weighted geometric mean of sand, silt and clay sizes. Prior studies (e.g., Koltermann and Gorelick (1995) and Porter et al. (2013)) have noted that this assumption may not hold if finer particles partially fill voids among coarser ones.

Results of the Glover et al. (2006) model (Fig. 1c) exhibited a pattern similar to the Kozeny-Carman model, overestimating $K_{sat}$ for smaller values and underestimating it for larger ones. This similarity most probably reflects the shared dependence on $d_a$ approximated from textural fractions. The RGPZ model produced results highly correlated ($R^2$ = 0.97) with those of the Kozeny-Carman model but, on average, estimated values one order of magnitude smaller (results not shown), with RMSLE = 1.28. Thus, while the RGPZ model performed better than the Kozeny-Carman model, its accuracy remained below that of CPA. This difference may partly stem from CPA's use of measured soil water retention data and estimated formation factors, whereas the other models rely solely on texture-based estimates.

For the Johnson et al. (1986) model (Eq. 5), RMSLE = 1.18 (Fig. 1d), indicating a general tendency to overestimate $K_{sat}$. Although both this model and CPA are based on the pore-throat diameter at the inflection point of the water retention curve, the performance difference may be linked to their constant coefficients – 8 for the Johnson et al. model and 53.5 for CPA. The sensitivity of results to this constant introduces additional uncertainty, as Schwartz and Banavar (1989) found that a smaller value (4) produced more accurate predictions for mono-sized granular composites.

The Mishra and Parker (1990) model substantially overestimated $K_{sat}$ (RMSLE = 2.68; Fig. 1e), consistent with observations by Pollacco et al. (2013) that this model tends to underestimate low conductivities and overestimate high ones. Similarly, the Guarracino (2007) model (Eq. 7) yielded RMSLE = 1.40 (Fig. 1f), comparable to the Kozeny-Carman and RGPZ models. Both models derive $K_{sat}$ from the van Genuchten (1980) parameters rather than directly from textural or structural information, which may partly explain their systematic deviations.

## 4. DISCUSSION

Based on the RMSLE values, the CPA model (Eq. 1 in combination with Eq. 2) yielded the most accurate overall estimates among the approaches evaluated. However, this outcome should be interpreted with caution. The CPA framework incorporates several simplifying assumptions, including the use of a fixed crossover porosity ($\phi_x$). Although early studies (e.g., Ghanbarian and Hunt, 2014) suggested that $\phi_x = 0.75$ is a reasonable approximation for soils, later work (Ghanbarian et al., 2015b) demonstrated that this parameter is not universal and may vary among soils depending on the breadth of their pore size distributions. We examined the sensitivity of the CPA approach to the choice of crossover porosity ($\phi_x$) in estimating $K_{sat}$. Specifically, we tested $\phi_x = 0.65$ and $\phi_x = 0.85$ in combination with $\phi_c = \theta_r$ to first calculate $F$ using Eq. (2) and subsequently estimate $K_{sat}$ using Eq. (1). The results of this sensitivity analysis are presented in Figs. 2a and 2b. The obtained RMSLE values were 0.88 for both $\phi_x = 0.65$ and 0.85, which are nearly identical to the RMSLE = 0.87 reported in Fig. 1a. This finding suggests that $K_{sat}$ estimates are not highly sensitive to variations in the choice of $\phi_x$. Figure 2c shows results based on $F$ derived from Archie's law (Archie, 1942), where Eq. (2) was applied with $\phi_x = 1$ and $\phi_c = 0$. The resulting RMSLE of 0.89 further indicates that CPA model performance is more strongly influenced by the accuracy of $d_c$ estimation in Eq. (1) than by the formulation of $F$.

Furthermore, the CPA-based approach relies on a single characteristic pore size corresponding to the inflection point of the soil water retention curve. While this assumption is suitable for soils with unimodal pore size distributions, it may not adequately represent soils exhibiting multimodal structures, for which our current understanding remains limited. In addition, the CPA dependence on the measured soil water retention curve constrains its applicability in data-scarce contexts. Therefore, although the CPA shows strong potential for capturing flow behavior in fine-textured soils, further validation using more diverse datasets and soil types are essential to assess its broader generality and robustness.

Although CPA offers a physically based and computationally consistent framework, it relies on accurate measurements of the soil water retention curve, which are more labor- and time-intensive than obtaining basic soil properties (e.g., texture or bulk density) used in empirical pedotransfer functions. This reliance is not unique to CPA, as models such as Mishra and Parker (1990) and Guarracino (2007) also require retention data. These models were primarily developed for research applications where water retention measurements are available. Thus, the advantage of CPA lies not in its practicality for large-scale applications but in its mechanistic representation of pore connectivity and transport processes once the required data are obtained.

Although the theoretical formulation of the CPA model is not new, its application to soil hydraulic characterization, particularly for estimating $K_{sat}$, has been very limited. The present study contributes to filling this gap by testing and validating the CPA-based approach on a large, field-scale dataset from the Kansas Mesonet, which is predominantly fine-textured (Table 1). Accordingly, the results should be viewed as representative of Kansas conditions rather than universally generalizable. The objective was to evaluate the performance of CPA in fine-textured soils, which are typically challenging for hydraulic conductivity estimation due to their complex pore networks. The promising performance of CPA in this dataset highlights its potential applicability to similar soil types, but comprehensive validation using soils of varied textures, structures, and climatic settings is necessary before broader generalization.

## 4. CONCLUSIONS

This study compared six models of saturated hydraulic conductivity, $K_{sat}$, using a database of 313 undisturbed soil samples. Our analysis showed that the critical path analysis (CPA) approach provided a robust theoretical framework to estimate $K_{sat}$ with a root mean square log-transformed error (RMSLE) of 0.87, outperforming the Kozeny-Carman, RGPZ, Johnson et al., Mishra and Parker, and Guarracino models. We found most predictions (81%) by the CPA model to be within one order of magnitude of the measured values, suggesting that the CPA model can capture the dominant transport pathways in undisturbed soil cores. The superior performance of CPA is likely attributable to its incorporation of pore-scale characteristics derived from the soil water retention curve information and its theoretical basis. In contrast, the alternative and more parsimonious models exhibited notable biases, particularly at very low (<10 cm day$^{-1}$) and very high (>10 cm day$^{-1}$) hydraulic conductivity values. The additional sensitivity analysis performed here revealed that the CPA predictions are relatively insensitive to variations in $\phi_x$ within a reasonable range (0.65–0.85), with RMSLE values showing minimal change. This finding demonstrates the robustness of the CPA approach and underscores its potential for broader use in soil physics. Given that most existing empirical models were calibrated on coarse-textured soils, the strong performance of the CPA in fine-textured Kansas soils provides new insights into its applicability and highlights its promise for extending physically based modeling to a wider spectrum of soil types.

**Acknowledgement**

BG is grateful to the University of Texas at Arlington for financial supports through the faculty startup fund and STARs award.

# REFERENCES

Ahmadisharaf, A., Nematirad, R., Sabouri, S., Pachepsky, Y., & Ghanbarian, B. (2024). Representative sample size for estimating saturated hydraulic conductivity via machine learning: A proof-of-concept study. Water Resources Research, 60(8), e2023WR036783.

Archie, G. E. (1942). The electrical resistivity log as an aid in determining some reservoir characteristics. Transactions of the AIME, 146(01), 54-62.

Bittelli, M., Campbell, G. S., & Flury, M. (1999). Characterization of particle-size distribution in soils with a fragmentation model. Soil Science Society of America Journal, 63(4), 782-788.

Carman, P. C. (1937). Fluid flow through granular beds. Trans. Inst. Chem. Eng. London, 15, 150-156.

Carsel, R. F., & Parrish, R. S. (1988). Developing joint probability distributions of soil water retention characteristics. Water Resources Research, 24(5), 755-769.

Chapuis, R. P. (2012). Predicting the saturated hydraulic conductivity of soils: a review. Bulletin of Engineering Geology and the Environment, 71, 401-434.

Chapuis, R. P., & Aubertin, M. (2003). On the use of the Kozeny Carman equation to predict the hydraulic conductivity of soils. Canadian Geotechnical Journal, 40(3), 616-628.

Dane, J.H. & Hopmans, J.W. (2002). Laboratory. Methods of Soil Analysis: Part 4 Physical Methods, 5, 675-720.

Doyen, P. M. (1988). Permeability, conductivity, and pore geometry of sandstone. Journal of Geophysical Research: Solid Earth, 93(B7), 7729-7740.

Dullien, F.A.L. (1992). Porous Media: Fluid Transport and Pore Structure. Academic Press, San Diego, CA.

Esmaeilpour, M., Ghanbarian, B., Sousa, R., & King, P. R. (2023). Estimating permeability and its scale dependence at pore scale using renormalization group theory. Water Resources Research, 59(5), e2022WR033462.

Guarracino, L. (2007). Estimation of saturated hydraulic conductivity Ks from the van Genuchten shape parameter α. Water resources research, 43(11). W11502.

Ghanbarian, B. (2021). Predicting Single-Phase Permeability of Porous Media Using Critical-Path Analysis. Complex Media and Percolation Theory, 273-288.

Ghanbarian, B., & Hunt, A. G. (2014). Universal scaling of gas diffusion in porous media. Water Resources Research, 50(3), 2242-2256.

Ghanbarian, B., & Skaggs, T. H. (2022). Soil water retention curve inflection point: Insight into soil structure from percolation theory. Soil Science Society of America Journal, 86(2), 338-344.

Ghanbarian, B., & Hunt, A. G. (2014). Universal scaling of gas diffusion in porous media. Water Resources Research, 50(3), 2242-2256.

Ghanbarian, B., Hunt, A. G., Ewing, R. P., & Skinner, T. E. (2014). Universal scaling of the formation factor in porous media derived by combining percolation and effective medium theories. Geophysical Research Letters, 41(11), 3884-3890.

Ghanbarian, B., Taslimitehrani, V., Dong, G., & Pachepsky, Y. A. (2015a). Sample dimensions effect on prediction of soil water retention curve and saturated hydraulic conductivity. Journal of Hydrology, 528, 127-137.

Ghanbarian, B., Daigle, H., Hunt, A. G., Ewing, R. P., & Sahimi, M. (2015b). Gas and solute diffusion in partially saturated porous media: Percolation theory and effective medium approximation compared with lattice Boltzmann simulations. Journal of Geophysical Research: Solid Earth, 120(1), 182-190.

Ghanbarian, B., Torres-Verdín, C., & Skaggs, T. H. (2016). Quantifying tight-gas sandstone permeability via critical path analysis. Advances in Water Resources, 92, 316-322.

Ghanbarian, B., Torres-Verdín, C., Lake, L. W., & Marder, M. (2019a). Gas permeability in unconventional tight sandstones: Scaling up from pore to core. Journal of Petroleum Science and Engineering, 173, 1163-1172.

Ghanbarian, B., Lake, L. W., & Sahimi, M. (2019b). Insights into rock typing: a critical study. SPE Journal, 24(01), 230-242.

Glover, P. W., Zadjali, I. I., & Frew, K. A. (2006). Permeability prediction from MICP and NMR data using an electrokinetic approach. Geophysics, 71(4), F49-F60.

Hunt, A. G. (2001). Applications of percolation theory to porous media with distributed local conductances. Advances in Water Resources, 24(3-4), 279-307.

Hunt, A. G. (2004). Percolative transport in fractal porous media. Chaos, Solitons & Fractals, 19(2), 309-325.

Hunt, A. G., & Sahimi, M. (2017). Flow, transport, and reaction in porous media: Percolation scaling, critical-path analysis, and effective medium approximation. Reviews of Geophysics, 55(4), 993-1078.

Hunt, A., Ewing, R., & Ghanbarian, B. (2014). Percolation theory for flow in porous media (Vol. 880). Springer.

Karim, M. R., & Krabbenhoft, K. (2010). New renormalization schemes for conductivity upscaling in heterogeneous media. Transport in Porous Media, 85, 677-690.

Katz, A. J., & Thompson, A. H. (1986). Quantitative prediction of permeability in porous rock. Physical Review B, 34(11), 8179.

Klute, A., & Dirksen, C. (1986). Hydraulic conductivity and diffusivity: Laboratory methods. In A. Klute (Ed.), Methods of soil analysis: Part 1 Physical and mineralogical methods (vol. 5, pp. 687–734). SSSA.

Koltermann, C. E., & Gorelick, S. M. (1995). Fractional packing model for hydraulic conductivity derived from sediment mixtures. Water Resources Research, 31(12), 3283-3297.

Kozeny, J. (1927). Ueber kapillare leitung des wassers im boden. Sitzungsberichte der Akademie der Wissenschaften in Wien, 136, 271-306.

Lin, H. S., McInnes, K. J., Wilding, L. P., & Hallmark, C. T. (1999). Effects of soil morphology on hydraulic properties II. Hydraulic pedotransfer functions. Soil Science Society of America Journal, 63(4), 955-961.

Mishra, S., & Parker, J. C. (1990). On the relation between saturated conductivity and capillary retention characteristics. Groundwater, 28(5), 775-777.

Parker, N., Kluitenberg, G. J., Redmond, C., & Patrignani, A. (2022). A database of soil physical properties for the Kansas Mesonet. Soil Science Society of America Journal, 86(6), 1495-1508.

Patrignani, A., Knapp, M., Redmond, C., & Santos, E. (2020). Technical overview of the Kansas Mesonet. Journal of Atmospheric and Oceanic Technology, 37(12), 2167-2183.

Peters, A., Hohenbrink, T. L., Iden, S. C., van Genuchten, M. T., & Durner, W. (2023). Prediction of the absolute hydraulic conductivity function from soil water retention data. Hydrology and Earth System Sciences Discussions, 2023, 1-32.

Pollacco, J. A. P., Nasta, P., Soria-Ugalde, J. M., Angulo-Jaramillo, R., Lassabatere, L., Mohanty, B. P., & Romano, N. (2013). Reduction of feasible parameter space of the inverted soil hydraulic parameter sets for Kosugi model. Soil Science, 178(6), 267-280.

Porter, L. B., Ritzi, R. W., Mastera, L. J., Dominic, D. F., & Ghanbarian-Alavijeh, B. (2013). The Kozeny-Carman equation with a percolation threshold. Groundwater, 51(1), 92-99.

Prosperini, N., & Perugini, D. (2008). Particle size distributions of some soils from the Umbria Region (Italy): Fractal analysis and numerical modelling. Geoderma, 145(3-4), 185-195.

Raats, P. A. C., & Klute, A. (1968). Transport in soils: The balance of mass. Soil Science Society of America Journal, 32(2), 161-166.

Regalado, C. M., & Muñoz-Carpena, R. (2004). Estimating the saturated hydraulic conductivity in a spatially variable soil with different permeameters: a stochastic Kozeny–Carman relation. Soil and Tillage Research, 77(2), 189-202.

Reynolds, W.D. & Elrick, D.E. (2002). Constant head well permeameter (vadose zone). Methods of Soil Analysis: Part 4 Physical Methods, 5: 844-858.

Roustazadeh, A., Ghanbarian, B., Shadmand, M. B., Taslimitehrani, V., & Lake, L. W. (2024). Estimating hydrocarbon recovery factor at reservoir scale via machine learning: Database-dependent accuracy and reliability. Engineering Applications of Artificial Intelligence, 128, 107500.

Sahimi, M. (2023). Applications of Percolation Theory. Springer. Second Edition. pp 679.

Schaap, M. G., & Leij, F. J. (1998). Database-related accuracy and uncertainty of pedotransfer functions. Soil Science, 163(10), 765-779.

Schaap, M. G., Leij, F. J., & Van Genuchten, M. T. (2001). Rosetta: A computer program for estimating soil hydraulic parameters with hierarchical pedotransfer functions. Journal of hydrology, 251(3-4), 163-176.

Schwartz, L. M., & Banavar, J. R. (1989). Transport properties of disordered continuum systems. Physical Review B, 39(16), 11965.

Singh, P. N., & Wallender, W. W. (2008). Effects of adsorbed water layer in predicting saturated hydraulic conductivity for clays with Kozeny–Carman equation. Journal of Geotechnical and Geoenvironmental Engineering, 134(6), 829-836.

Skaggs, T. H. (2011). Assessment of critical path analyses of the relationship between permeability and electrical conductivity of pore networks. Advances in Water Resources, 34(10), 1335-1342.

Sullivan, P. L., Billings, S. A., Hirmas, D., Li, L., Zhang, X., Ziegler, S., ... & Wen, H. (2022). Embracing the dynamic nature of soil structure: A paradigm illuminating the role of life in critical zones of the Anthropocene. Earth-Science Reviews, 225, 103873.

Tietje, O., & Hennings, V. (1996). Accuracy of the saturated hydraulic conductivity prediction by pedo-transfer functions compared to the variability within FAO textural classes. Geoderma, 69(1-2), 71-84.

van Genuchten, M. T. (1980). A closed-form equation for predicting the hydraulic conductivity of unsaturated soils. Soil Science Society of America Journal, 44(5), 892-898.

Van Looy, K., Bouma, J., Herbst, M., Koestel, J., Minasny, B., Mishra, U., Montzka, C., Nemes, A., Pachepsky, Y. A., Padarian, J., Schaap, M. G., Tóth, B., Verhoef, A., Vanderborght, J., van der Ploeg, M. J., & Vereecken, H. (2017). Pedotransfer functions in Earth system science: Challenges and perspectives. Reviews of Geophysics, 55, 1199–1256.

Weber, T. K. D., Weihermüller, L., Nemes, A., Bechtold, M., Degré, A., Diamantopoulos, E., ... & Bonetti, S. (2024). Hydro-pedotransfer functions: a roadmap for future development. Hydrology and Earth System Sciences, 28(14), 3391-3433.

Wu, Q., Borkovec, M., & Sticher, H. (1993). On particle-size distributions in soils. Soil Science Society of America Journal, 57(4), 883-890.

Xu, P., & Yu, B. (2008). Developing a new form of permeability and Kozeny–Carman constant for homogeneous porous media by means of fractal geometry. Advances in Water Resources, 31(1), 74-81.

Yong, L., Chengmin, H., Baoliang, W., Xiafei, T., & Jingjing, L. (2017). A unified expression for grain size distribution of soils. Geoderma, 288, 105-119.

Zhang, Y., & Schaap, M. G. (2017). Weighted recalibration of the Rosetta pedotransfer model with improved estimates of hydraulic parameter distributions and summary statistics (Rosetta3). Journal of Hydrology, 547, 39-53.

Zhang, Y., & Schaap, M. G. (2019). Estimation of saturated hydraulic conductivity with pedotransfer functions: A review. Journal of Hydrology, 575, 1011-1030.

**Table 1**. Median values for selected physical properties of the Kansas Mesonet soil database.

| Textural class | N | Sand | Clay | Porosity | Bulk density | $K_{sat}$ |
|---|---|---|---|---|---|---|
| | | % | % | $cm^3\ cm^{-3}$ | $g\ cm^{-3}$ | $cm\ day^{-1}$ |
| Clay | 10 | 5 | 56 | 0.48 | 1.38 | 0.5 |
| Clay loam | 36 | 25 | 32 | 0.46 | 1.44 | 18.7 |
| Loam | 16 | 37 | 24 | 0.465 | 1.46 | 41 |
| Sandy clay loam | 2 | 48 | 25 | 0.415 | 1.55 | 0.45 |
| Sandy loam | 24 | 65 | 13 | 0.38 | 1.65 | 47.1 |
| Silt loam | 72 | 16 | 24 | 0.475 | 1.35 | 13.9 |
| Silty clay | 50 | 6 | 45 | 0.49 | 1.36 | 1.5 |
| Silty clay loam | 103 | 10 | 32 | 0.49 | 1.34 | 32.1 |

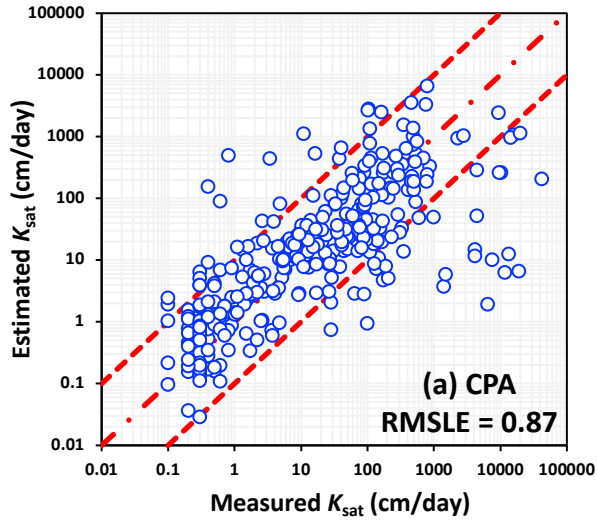
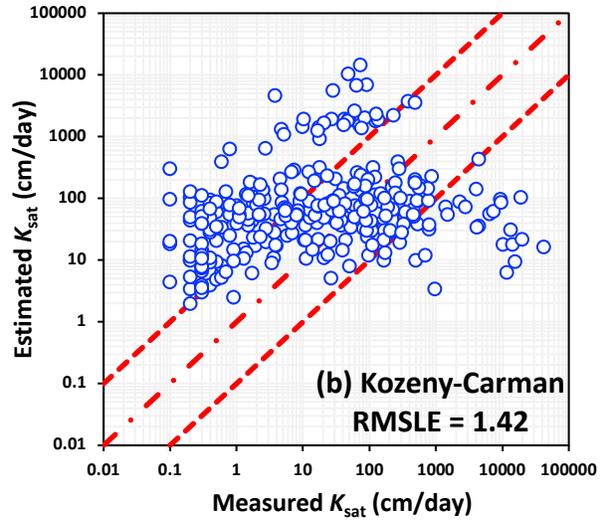
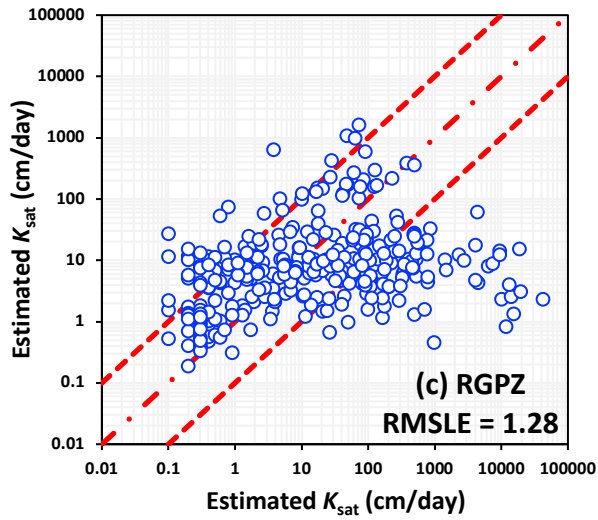
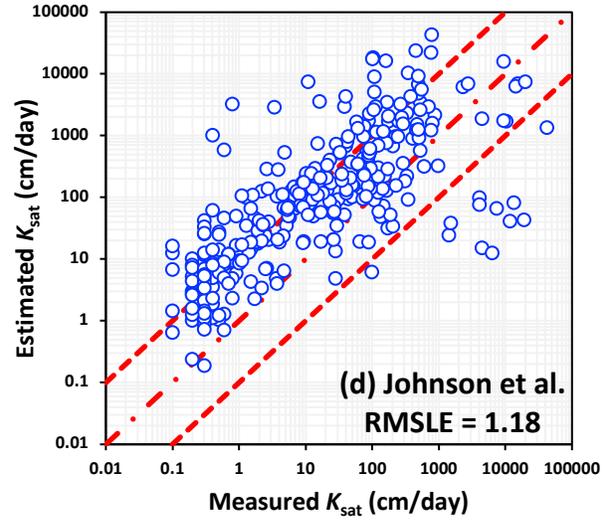
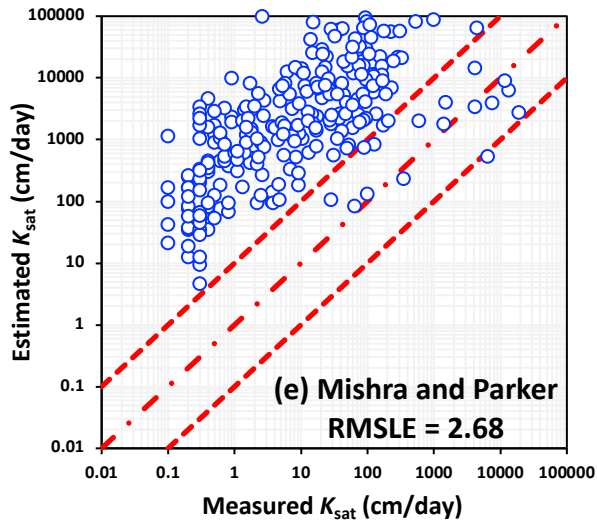
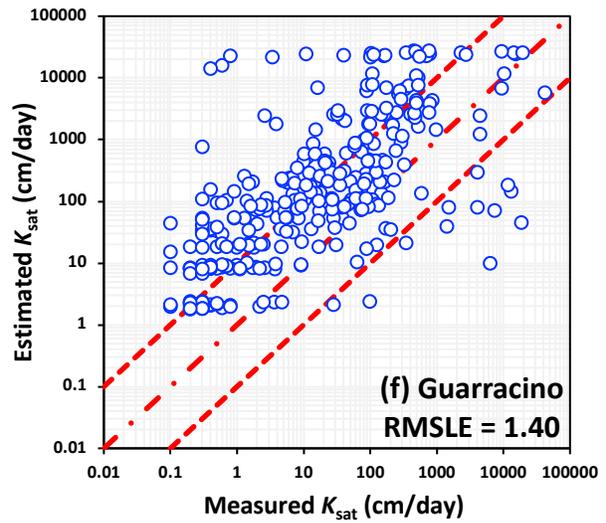

**Figure 1**. Estimated saturated hydraulic conductivity using (a) critical path analysis, Eq. (1), (b) Kozeny-Carman model, Eq. (3), (c) Glover et al. (2006) or RGPZ model, Eq. (4), (d) Johnson et al. (1986) model, Eq. (5), (e) Mishra and Parker (1990) model, Eq. (6), and (f) Guarracino (2007) model, Eq. (7), versus the measured value for 313 soil samples from the Kansas Mesonet database.

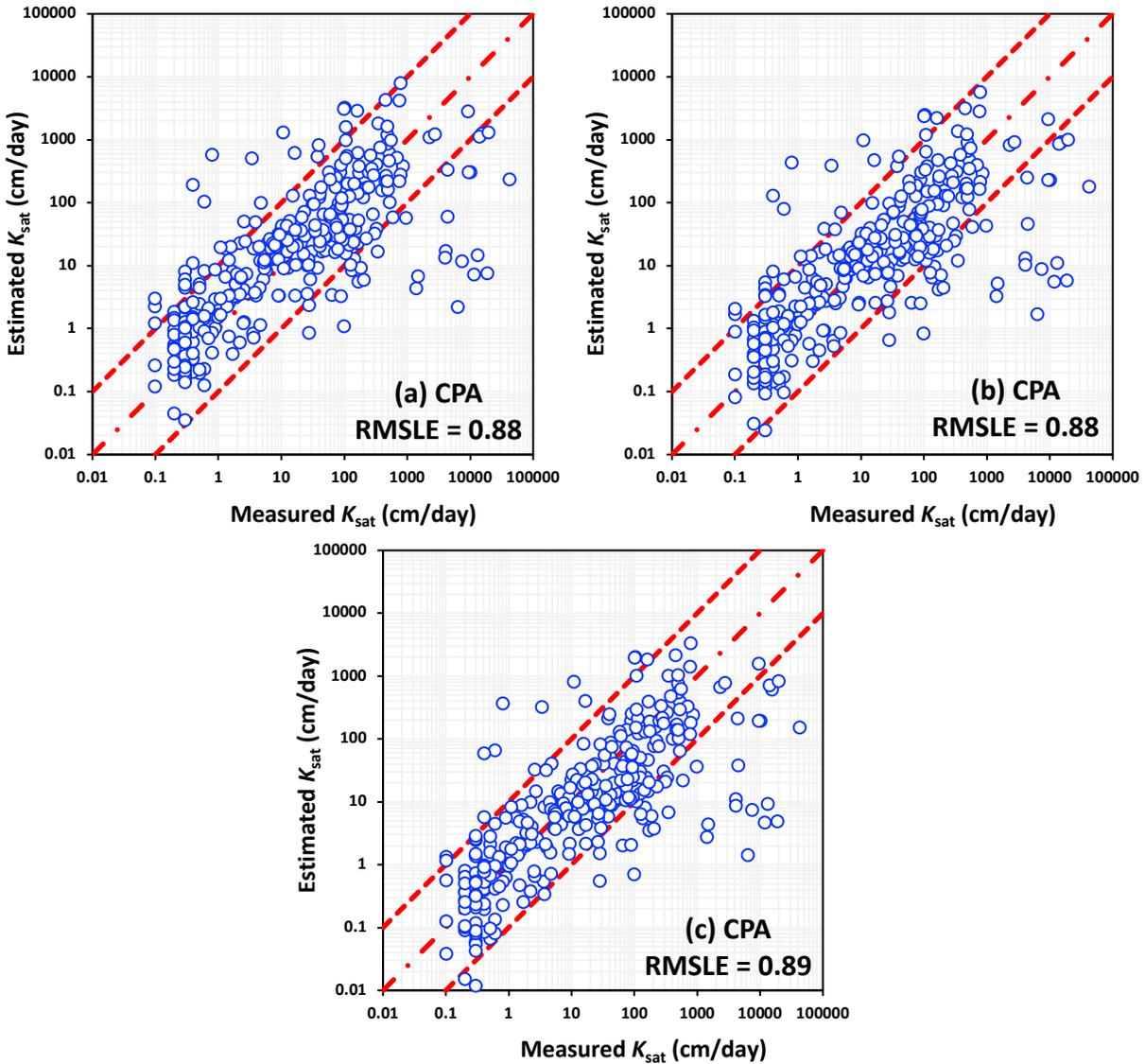

Figure 2. Estimated saturated hydraulic conductivity via the critical path analysis, Eq. (1), using $\phi_x = 0.65$ (a) and 0.85 (b) in Eq. (2) versus the measured value for 313 soil samples from the Kansas Mesonet database. (c) shows the estimated saturated hydraulic conductivity via the critical path analysis, Eq. (1), using $\phi_x = 1$ and $\phi_c = 0$, consistent with Archie's law (Archie, 1942).